\documentclass[reprint,superscriptaddress,preprintnumbers,amsmath,amssymb,prb,showkeys,longbibliography,floatfix,balancelastpage]{revtex4-2}
\usepackage{graphicx}
\usepackage{xcolor}
\usepackage[unicode=true, colorlinks=true,linktoc=all,pdfusetitle, linkcolor=black, citecolor=black, urlcolor=black,final]{hyperref}
\usepackage[separate-uncertainty=true]{siunitx}
\begin{document}

\title{Spectral fingerprint of quantum confinement in single \texorpdfstring{CsPbBr\textsubscript{3}}{CsPbBr3} nanocrystals}%

\author{Mohamed-Raouf Amara}
 
\affiliation{%
 Laboratoire de Physique de l'\'Ecole Normale Sup\'erieure, ENS, Université PSL, CNRS, Sorbonne Universit\'e, Universit\'e Paris-Cité, F-75005 Paris, France
}%
\affiliation{%
 Division of Physics and Applied Physics, School of Physical and Mathematical Sciences, Nanyang Technological University, 637371, Singapore
}%

\author{Zakaria Said}
\affiliation{%
 Laboratoire de Physique de l'\'Ecole Normale Sup\'erieure, ENS, Université PSL, CNRS, Sorbonne Universit\'e, Universit\'e Paris-Cité, F-75005 Paris, France
}%
\author{Caixia Huo}%
\affiliation{%
 Division of Physics and Applied Physics, School of Physical and Mathematical Sciences, Nanyang Technological University, 637371, Singapore
}%
\author{Aurélie Pierret}
\author{Christophe Voisin}
\affiliation{%
 Laboratoire de Physique de l'\'Ecole Normale Sup\'erieure, ENS, Université PSL, CNRS, Sorbonne Universit\'e, Universit\'e Paris-Cité, F-75005 Paris, France
}%
\author{Weibo Gao}
\affiliation{%
 Division of Physics and Applied Physics, School of Physical and Mathematical Sciences, Nanyang Technological University, 637371, Singapore
}%
\author{Qihua Xiong}
\altaffiliation{%
Current affiliation:
State Key Laboratory of Low-Dimensional Quantum Physics, Department of Physics, Tsinghua University, Beijing 100084, China.
}%
\altaffiliation{%
Frontier Science Center for Quantum Information, Beijing 100084, P.R. China
}%
\altaffiliation{%
Collaborative Innovation Center of Quantum Matter, Beijing 100084, P.R. China}%
\altaffiliation{%
Beijing Academy of Quantum Information Sciences, Beijing 100193, P.R. China}
\affiliation{%
 Division of Physics and Applied Physics, School of Physical and Mathematical Sciences, Nanyang Technological University, 637371, Singapore
}%
\author{Carole Diederichs}
\email{Corresponding author: carole.diederichs@phys.ens.fr}
\affiliation{%
 Laboratoire de Physique de l'\'Ecole Normale Sup\'erieure, ENS, Université PSL, CNRS, Sorbonne Universit\'e, Universit\'e Paris-Cité, F-75005 Paris, France
}%
\affiliation{%
 Institut Universitaire de France (IUF), 75231 Paris, France
}%

\date{March 30, 2023}%
\begin{abstract}
\begin{center} \textbf{Abstract} \end{center}
{Lead halide perovskite nanocrystals are promising materials for classical and quantum light emission. To understand these outstanding properties, a thorough analysis of the band-edge exciton emission is needed which is not reachable in ensemble and room temperature studies because of broadening effects. %
Here, we report on a cryogenic-temperature study of the photoluminescence of single CsPbBr\textsubscript{3} NCs in the intermediate quantum confinement regime. We reveal the size-dependence of the spectral features observed: the bright-triplet exciton energy splittings, the trion and biexciton binding energies as well as the optical phonon replica spectrum. In addition, we show that bright triplet energy splittings are consistent with a pure exchange model and that the variety of polarisation properties and spectra recorded can be rationalised simply by considering the orientation of the emitting dipoles and the populations of the emitting states.}
\end{abstract}
\keywords{\texttt{single perovskite nanocrystals, photoluminescence, excitonic complexes, phonon replica}}
\maketitle

Lead halide perovskite nanocrystals (LHP NCs) have gained significant interest for both photovoltaic and light-emission applications due to their outstanding performance and intriguing optoelectronic properties~\cite{greenEmergencePerovskiteSolar2014,sutherlandPerovskitePhotonicSources2016}. %
Early reports of photoluminescence (PL) of individual LHP NCs have revealed emission of antibunched single photons across a wide range of temperatures~\cite{parkRoomTemperatureSinglePhoton2015,huSuperiorOpticalProperties2015,rainoSingleCesiumLead2016} together with short radiative lifetimes~\cite{rainoSingleCesiumLead2016,beckerBrightTripletExcitons2018}.  
In addition to this main emission channel attributed to the bright triplet exciton in LHP NCs~\cite{ramadeFineStructureExcitons2018a,fuNeutralChargedExciton2017,beckerBrightTripletExcitons2018}, the PL spectra of single LHP NCs also display a large number of additional emission peaks attributed to charged exciton and biexciton emissions as well as a set of phonon replica, all located within a 20 meV bandwidth~\cite{fuNeutralChargedExciton2017,yinBrightExcitonFineStructureSplittings2017,fuUnravelingExcitonPhonon2018,choLuminescenceFineStructures2021}. %
The contribution of these states to the room-temperature ensemble emission of NCs may be completely blurred because of both inhomogeneous and homogeneous broadening, such that these can only be resolved in low-temperature single NC experiments which can access the entire fine structure of the PL spectrum of LHP NCs.

In contrast with established NCs, usual LHP NCs have sizes that range from a few to many times the exciton Bohr radius. As such, LHP NCs are in an intermediate confinement regime where charges are weakly confined by the structure, and electron and hole motions are correlated through Coulomb interaction. 
As the optical properties of semiconductor NCs result from an interplay between the band-edge exciton states, their interaction with phonons and with the surface state of the NCs, a precise understanding of the band-edge exciton manifold emission and the vibronic replica is critical to assess the potential of LHP NCs as quantum light emitters and identify their peculiarities. In particular, the influence of the Rashba effect on the exciton fine structure of LHP NCs has been the subject of an intense debate in the community~\cite{beckerBrightTripletExcitons2018,benaichBrightExcitonSplittingsInorganic2019,benaichMultibandModelTetragonal2020,sercelExcitonFineStructure2019,sercelQuasicubicModelMetal2019}.

In this work, we provide a thorough study of the spectral properties of single CsPbBr\textsubscript{3} NCs at cryogenic temperature over a wide range of emission energies, corresponding to NCs between 5 and $\sim$\SI{20}{\nano\meter}. Based on a combination of scanning confocal microscopy, polarisation measurements and electron microscopy, we analyse the full set of spectral features related to the emission, namely the bright triplet exciton, trion, biexciton and their respective optical phonon replica, as a function of NC size. The variety of bright triplet exciton structures observed is rationalised as arising from a combination of the orientation of the emitting dipoles with respect to the detection axis and the underlying population of each emitting state. %
Bright triplet energy splittings are found to increase with decreasing NC size, in accordance with a pure exchange model. Similarly, the trion and biexciton binding energies increase with confinement.

CsPbBr\textsubscript{3} NCs were synthesised following the hot-injection method using a slightly modified synthesis~\cite{beckerBrightTripletExcitons2018} (see SIa). Notably, the injection temperature was used to control the size of the four batches of NCs studied. %
Electron microscopy reveals NCs sizes from 5 to $\sim$~\SI{20}{\nano\meter} with average aspect ratios $\sim$~1.1. %
The NCs edge lengths distributions are compared to the distributions of single NC PL emission energies to obtain the NC size-emission energy relationship (see SIc). %
For single NC spectroscopy, solutions diluted in a mixture of toluene and polystyrene are spin-coated on dielectric substrates and studied using a home-built scanning confocal microscope based on a 0.7 numerical aperture objective (see SIa,b).

\begin{figure}
\includegraphics[]{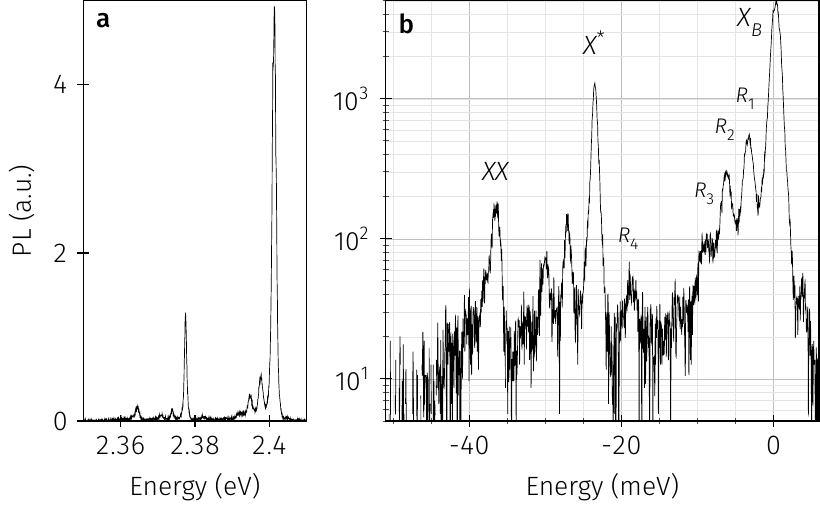}
\caption{Exemplary spectrum of a single CsPbBr\textsubscript{3} NC shown in (a) linear and (b) logarithmic scale. The emission spectrum displays signatures of the bright exciton $X_B$, trion $X^{\text{*}}$, and biexciton $XX$ together with their respective optical phonon replica, labelled $R_i$ for $X_B$.}
\label{fig:spectre}
\end{figure}

The PL spectrum of a single LHP NC at cryogenic temperature displays a variety of spectral signatures (\autoref{fig:spectre}) related to the bright exciton, trion and biexciton together with their respective optical phonon replica which are discussed sequentially hereafter.

\begin{figure*}[bt]
\includegraphics[]{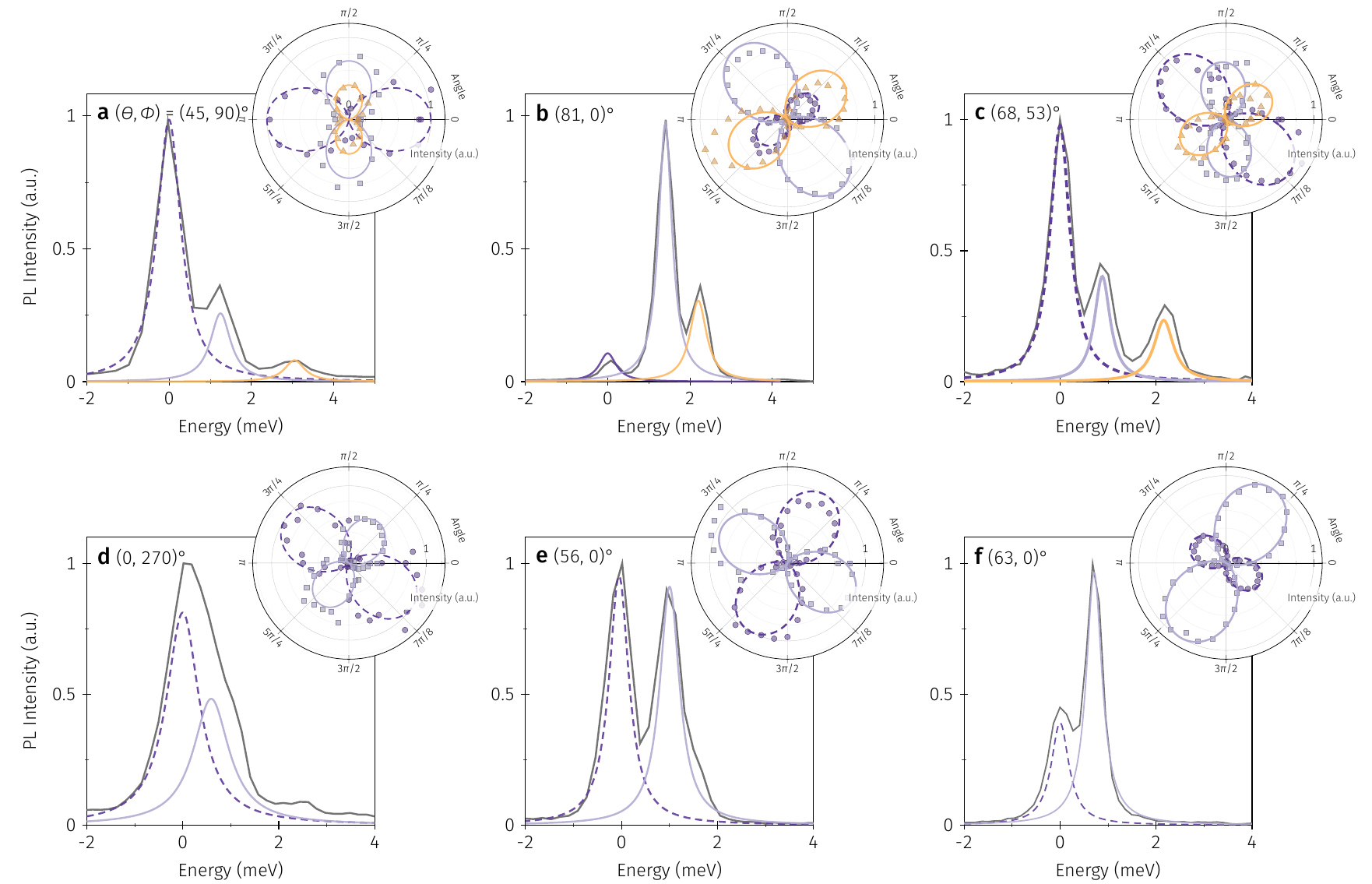}%
\caption{Polarisation of the bright triplet exciton in single CsPbBr$_3$ NCs. Spectrum and polarisation diagram of single NCs with (a-c) three peaks or (d-f) two peaks attributed to the bright exciton. (dark line) Experimental spectrum obtained as a sum of all polarisations, (colors) simulated spectrum components. Insets: (dots) experimental intensity of each emission line as a function of the detection angle, (lines) simulated intensity. Observation directions of the three orthogonal dipoles are given in each panel as $(\Theta, \Phi)$. Relative intensities of emission lines in insets are adjusted for clarity whenever necessary.}
\label{fig:polar}
\end{figure*}

First, the bright exciton emission consists of either two or three peaks (\autoref{fig:polar}), consistent with prior studies at cryogenic temperature~\cite{rainoSingleCesiumLead2016,isarovRashbaEffectSingle2017,yinBrightExcitonFineStructureSplittings2017,fuNeutralChargedExciton2017,ramadeFineStructureExcitons2018a,beckerBrightTripletExcitons2018}. %
The band-edge exciton of LHP NCs is formed by a Pb $p$-orbital electron and an $s$-like hole and is split by electron-hole exchange interaction into a dark singlet and a bright triplet~\cite{evenImportanceSpinOrbit2013,evenPedestrianGuideSymmetry2015}.
Depending on the actual symmetry of the NC, varying from apparently cubic at high temperature to orthorhombic at the lowest cryogenic temperatures, the bright triplet is expected to split into up to three non-degenerate states. In addition, the NC shape anisotropy has a similar effect whereby even a NC with cubic symmetry can exhibit a fully split triplet exciton in the presence of anisotropy~\cite{nestoklonOpticalOrientationAlignment2018a,benaichBrightExcitonSplittingsInorganic2019}. %
For CsPbBr\textsubscript{3} NCs, a cubic crystal structure was initially incorrectly reported~\cite{protesescuNanocrystalsCesiumLead2015a}, an assignment later attributed to the dynamic character of the perovskite lattice and revised to an orthorhombic crystal structure~\cite{cottinghamCrystalStructureColloidally2016,bertolottiCoherentNanotwinsDynamic2017,kovalenkoPropertiesPotentialOptoelectronic2017}. %
At the single NC level, the presence of two- and three-line spectra has been attributed respectively to NCs with tetragonal and orthorhombic crystal phase~\cite{fuNeutralChargedExciton2017,ramadeFineStructureExcitons2018a}.

The crystal structure of our isolated NCs at cryogenic temperature was not characterised, so we can only postulate on it using spectroscopic arguments related to the polarisation properties recorded. %
Besides, as shown in the following, even if all NCs have an orthorhombic crystal structure, two-peak spectra can still be observed which urges caution to a direct identification of a crystal structure based solely on the number of observable bright exciton peaks, especially in the absence of a magnetic field.

Here, we measure the relative intensities of the bright triplet exciton peaks as a function of the polarisation (\autoref{fig:polar}a-f). %
The majority of NCs exhibit two-peak spectra as those shown in~\autoref{fig:polar}d-f, characterised by linear and orthogonal polarisations. The relative intensities of the two peaks vary, from a dimmest to a brightest lowest bright exciton state, always with orthogonal polarisations.
NCs with a three-peak spectrum also exhibit linear polarisations with strong degrees of polarisation. In~\autoref{fig:polar}a-c, we show three configurations observed. While \autoref{fig:polar}a follows an intensity ordering in qualitative agreement with an expected thermal population, \autoref{fig:polar}b exhibits a dimmest lowest-exciton state. In both cases, we find two peaks with close to parallel polarisations while in~\autoref{fig:polar}c, each emission peak exhibits a distinct polarisation axis.

The observation over a large number of NCs of strong linear polarisations for all two-peak spectra, which would not necessarily be the case for NCs with cubic shape and tetragonal crystal structure, suggests that all spectra should be understood as arising from NCs with a fully split triplet. %
Further considering that our NCs have a narrow shape distribution with an aspect ratio close to 1 (see Figure S3) and that we probe a variety of NCs orientations in our samples, the absence of two-peak spectra with polarisation deviating from linear suggest that our NCs have a fully-split triplet regardless of morphology, i.e. an orthorhombic crystal structure. %
Regardless of its origin, be it crystal structure or morphology, in the following we assume a fully split triplet.

We model the fully-split bright triplet exciton as three orthogonal dipoles, along $\hat{e}_x$, $\hat{e}_y$, $\hat{e}_z$ (with $E_x<E_y<E_z$), with the same oscillator strength and investigate if the diversity of spectra can be explained as arising from a combination of the orientation of the observation axis with respect to the three dipoles and of the occupation of the emitting states. %
For our simulations, we fix the energy splittings to that observed experimentally, which sets the relative populations of the states via thermal statistics, and allow the observation direction to vary (see SII).
Doing so, we can determine the observation axis as defined by the light propagation vector $\vec{u} =i\hat{e_x}+j\hat{e_y}+k\hat{e_z}$ or given equivalently as the two spherical angles $\Theta$ and $\Phi$.

For \autoref{fig:polar}a, the spectrum is reproduced by assuming light propagation in the (yz) plane with $\Theta=\SI{45}{\degree}$. By adjusting the relative peak intensities, we find a temperature of \SI{15}{\kelvin} close to the actual temperature of \SI{10}{\kelvin}.  
A similar reasoning for \autoref{fig:polar}b,c yields close matches with an experimental temperature of \SI{10}{\kelvin} and observation directions given in \autoref{fig:polar}b,c. %
For two-peak spectra, the reason for the absence of the third exciton peak can be multifold.
For \autoref{fig:polar}d, we can assume that a third emission peak, with energy $E_k$, is not observed because its emission dipole is aligned with the observation direction. Doing so, we find a reasonable temperature of \SI{16}{\kelvin}. 
A similar reasoning for \autoref{fig:polar}e, with an observation direction parallel to one of the emission dipole, does not hold as a temperature of \SI{200}{\kelvin} would be needed to match the relative intensities recorded. For \autoref{fig:polar}e,f the absence of a third exciton peak at higher energy is rather interpreted as resulting from its low population. In these cases, observation directions given in \autoref{fig:polar}e,f yield a good match provided the third exciton peak is at least \SI{2}{\milli\electronvolt} above the observed doublet.

\begin{figure*}
\includegraphics[]{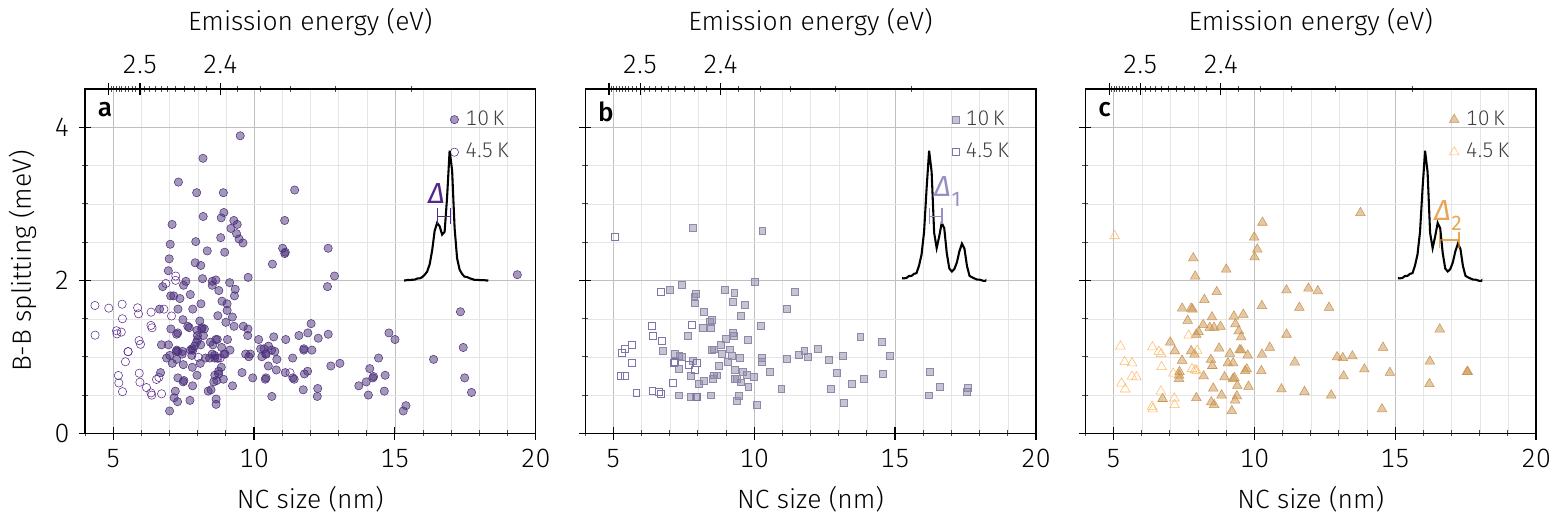}%
\caption{Size-dependence of bright-bright exciton energy splittings in single CsPbBr$_3$ NCs for: (a) two-peak spectra, (b-c) three-peak spectra. Filled (empty) markers correspond to data acquired at \SI{10}{\kelvin} (\SI{4.5}{\kelvin}). }
\label{fig:splittings}
\end{figure*}

Experimental spectra can thus adequately be reproduced by considering the combined effects of the emitting states populations and the observation direction with respect to the emitting dipoles. For orthorhombic NCs, the actual emission dipoles are expected to be parallel to the crystallographic axes~\cite{beckerBrightTripletExcitons2018}. In such case, provided the orientation of the crystallographic axes with respect to the NC morphological axes is known, as for nanoplatelets~\cite{bertolottiCrystalStructureMorphology2019,schmitzOpticalProbingCrystal2021}, the absolute orientation of NCs could be obtained (see Note SIIb). %
In any case, in our experiments, NCs are embedded in a $\sim$50-\SI{100}{\nano\meter} layer of polymer and are thus expected to exhibit random orientations which prohibits a definite assignment for the NCs orientations. 

The relative position of the singlet and triplet exciton states has been a subject of debate in the last years since the prediction that the bright triplet exciton could be the ground exciton state of LHP NCs due to Rashba effect~\cite{beckerBrightTripletExcitons2018}. %
While experimental results contradict this prediction~\cite{tamaratGroundExcitonState2019,tamaratDarkExcitonGround2020}, the discussion has now shifted towards quantifying the effect of this potential Rashba effect on the exciton fine structure. A point of interest in this discussion is the prediction related to bright-bright energy splittings.  %
While in a pure exchange model, bright-bright energy splittings are expected to increase with confinement~\cite{benaichBrightExcitonSplittingsInorganic2019}, including the Rashba effect yields the contradicting prediction~\cite{sercelExcitonFineStructure2019,sercelQuasicubicModelMetal2019}. %
Experimentally, studies on CsPbBr\textsubscript{3} NCs~\cite{ramadeFineStructureExcitons2018a} reveal no clear variation of energy splittings with NC size in contrast to CsPbI\textsubscript{3} NCs~\cite{tamaratDarkExcitonGround2020}, although similar ranges of NC sizes were considered.

Here, we investigate the energy splittings of the bright triplet states for a large range of NCs sizes (\autoref{fig:splittings}). %
Similar to earlier reports~\cite{rainoSingleCesiumLead2016,ramadeFineStructureExcitons2018a,beckerBrightTripletExcitons2018}, we find energy splittings on the order of \SI{1}{\milli\electronvolt} with no striking emission energy or size-dependence. Such dependence may be blurred by the dispersion in energy splittings at a given average NC size resulting from the distribution of NCs morphologies investigated (see Figure S3). %
For two-peak spectra, we find a mean energy splitting $\Delta = \SI{1.3\pm 0.7}{\milli\electronvolt}$ while for three-peak spectra we find $\Delta_1 = \SI{1.0\pm 0.5}{\milli\electronvolt}$ and $\Delta_2 = \SI{1.1\pm 0.6}{\milli\electronvolt}$ for the low and high energy splitting respectively. %
Considering that two-peak spectra can correspond to any pair of the bright triplet, the expected ratio between the average two-peak and three-peak energy splittings is 1.33~\cite{beckerBrightTripletExcitons2018}. Experimentally, we find $\bar{\Delta}_{2P} \sim 1.2 \bar{\Delta}_{3P}$, close to the expected value.

To reveal an underlying size-dependence, we average the observed energy splittings in two different size regimes: NCs larger and smaller than \SI{12}{\nano\meter} $\sim 4 a_B$. This threshold corresponds to a NC size below which the exchange parameter significantly deviates from its bulk value~\cite{sercelExcitonFineStructure2019,sercelQuasicubicModelMetal2019}. We can therefore calculate the ratio $r$ between energy splittings for small and large NCs, and get $r_{\Delta} \sim 1.35$, $r_{\Delta_1} \sim 1.42$ and $r_{\Delta_2} \sim 1.05$ with $r_{\Delta}$ consistently above 1, thus revealing a general increase of energy splittings with confinement. We also verified that these ratios are consistently above 1 independently of the chosen threshold between 10 and \SI{20}{\nano\meter}. %
This is in agreement with the predictions of a pure exchange model where bright-bright energy splittings are expected to increase from a bulk value of a few \SI{100}{\micro\electronvolt} to $\sim$~\SI{2}{\milli\electronvolt} for a \SI{5}{\nano\meter} size NC~\cite{benaichBrightExcitonSplittingsInorganic2019,sercelExcitonFineStructure2019}. %
The NCs shape should also contribute to setting bright-bright energy splittings even for small anisotropy factors. For CsPbI\textsubscript{3} NCs, this contribution was found on the order of a few \SI{100}{\micro\electronvolt} for large NCs and up to $\sim$~\SI{1}{\milli\electronvolt} for the smallest NCs~\cite{tamaratDarkExcitonGround2020}. 
Thus the statistical analysis of our results, although showing significant dispersion interpreted as a result of the slight shape anisotropy, reveal a dependence in agreement with the expected influence of the exchange interaction and do not show a signature related to the Rashba effect.

In addition to the bright, resp. dark, character of the triplet, resp. singlet, exciton, symmetry arguments can also be used to predict the fine structure of the two other excitonic complexes observed in our experiments.
As previously reported~\cite{yinBrightExcitonFineStructureSplittings2017,tamaratDarkExcitonGround2020}, the trion in LHP NCs does not have a fine structure while the biexciton exhibits a fine structure mirroring that of the bright exciton (\autoref{fig:spectre} and S6). At the low excitation powers used in this work, the biexciton emission intensity is weak. Notably, while the trion emission is intermittent and exhibits significant spectral jumps, the exciton and biexciton emission is more stable.

\begin{figure}[htbp]
\includegraphics[]{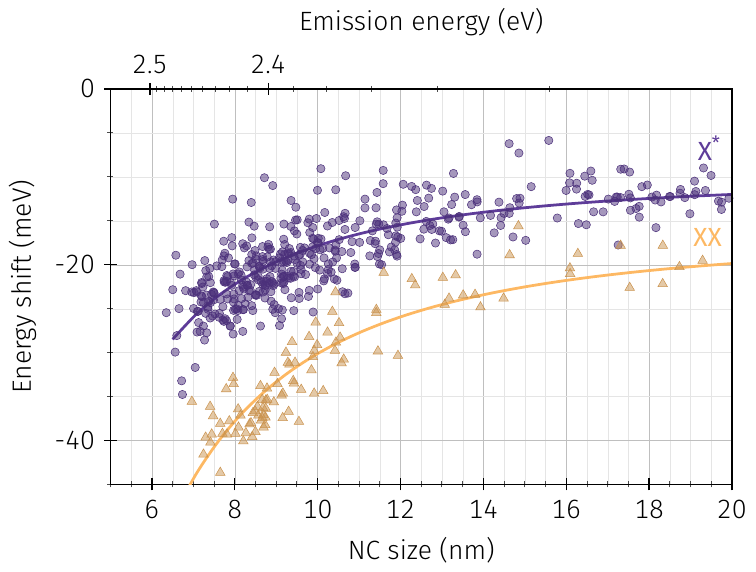}
\caption{Size-dependence of the trion (X\textsuperscript{*}) and biexciton (XX) binding energies in single \texorpdfstring{CsPbBr\textsubscript{3}}{CsPbBr3} NCs. Circles, resp. triangles, correspond to energy shifts recorded for the trion, resp. biexciton, with respect to the lowest neutral exciton line.}
\label{fig:trionbi}
\end{figure}

The trion and biexciton energy shifts with respect to the exciton show clear variations (\autoref{fig:trionbi}) between 10 and \SI{30}{\milli\electronvolt} for the trion and 15 and \SI{40}{\milli\electronvolt} for the biexciton. %
This dependency is similar to recent results on slightly larger-sized CsPbBr\textsubscript{3} NCs as a function of the emission energy which reported trion and biexciton binding energies up to \SI{16}{\milli\electronvolt} and \SI{35}{\milli\electronvolt} respectively~\cite{zhuManyBodyCorrelationsExciton2023,tamaratUniversalScalingLaws2023}.
To recover the bulk energy shifts and the size-dependence of the energy shifts, we use a simple empirical law of the form $A + B/L^2$. We find bulk energy shifts $A_{X\textsuperscript{*}} = \SI{-10}{\milli\electronvolt}$, $A_{XX} = \SI{-16.5}{\milli\electronvolt}$ and a size-dependence given by $B_{X\textsubscript{*}} = \SI{-780}{\milli\electronvolt\cdot\nano\meter\squared}$ and $B_{XX} = \SI{-1370}{\milli\electronvolt\cdot\nano\meter\squared}$. %
The biexciton shift shows good agreement with the size-dependence measured by two-dimensional electronic spectroscopy on ensembles of CsPbBr$_3$ NCs where $A_{XX} = \SI{-17.8}{\milli\electronvolt}$ and $B_{XX} = \SI{-1140}{\milli\electronvolt\cdot\nano\meter\squared}$~\cite{huangInhomogeneousBiexcitonBinding2020}. With such parameters, biexciton shifts up to \SI{100}{\milli\electronvolt} are expected for \SI{4}{\nano\meter}-sized NCs which is consistent with the largest reported values~\cite{castanedaEfficientBiexcitonInteraction2016}.

Redshifted phonon replica were observed at the single object level in all-inorganic LHP NCs~\cite{fuNeutralChargedExciton2017,tamaratDarkExcitonGround2020} where two modes are reported below 7 meV, and in hybrid LHP NCs~\cite{fuUnravelingExcitonPhonon2018,pfingstenPhononInteractionPhase2018,tamaratGroundExcitonState2019,choLuminescenceFineStructures2021} where up to three modes are identified with energies up to $\sim$~20 meV. %
Notably, for all-inorganic perovskites, while the main longitudinal optical phonon mode at $\sim$~16 meV is widely observed in Raman measurements~\cite{yaffeLocalPolarFluctuations2017,iaruFrohlichInteractionDominated2021} and recovered from temperature-dependent linewidth measurements~\cite{ramadeFineStructureExcitons2018a}, it was only recently observed in the low-temperature emission of single all-inorganic LHP NCs albeit at a higher energy~\cite{choExcitonPhononTrion2022a}.

\begin{figure}[htbp]
\includegraphics[]{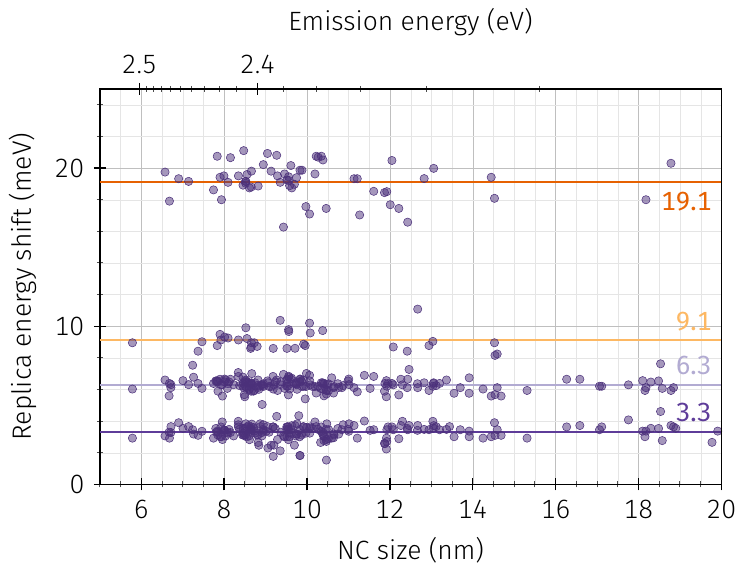}
\caption{Size-dependence of the four optical phonon replica associated to the bright triplet exciton in single \texorpdfstring{CsPbBr\textsubscript{3}}{CsPbBr3} NCs.}
\label{fig:replica}
\end{figure}

Here, we report on the observation of four distinct optical phonon replica in the low energy side of the exciton PL at 3.3, 6.3, 9.1 and \SI{19.1}{\milli\electronvolt} (\autoref{fig:replica} and S7). As summarised in~\autoref{fig:replica}, the energy of phonon replica relative to the bright exciton emission do not show any size-dependence. Two anti-Stokes optical phonon replica of the bright exciton are also observed for the brightest emitters with energies matching the lowest Stokes-shifted replica (see~Figure S5). Notably, the two, resp. three, lowest energy optical phonon modes also correspond to phonon replica observed for the biexciton, resp. trion. %

The absence of a size-dependence for the replica energies in such a large range of sizes, with surface to volume ratios varying by a factor of 5, is a clue to the origin of these modes. Confined acoustic modes and surface optical modes are expected to have a size-dependence related to the NC finite size and increased influence of the surface~\cite{cerulloSizedependentDynamicsCoherent1999,hwangSizedependentSurfacePhonon1999}. Our results thus indicate that the four replica observed in the PL of single NCs are related to bulk optical phonon modes of the orthorhombic perovskite lattice. %
While the replicas energies are independent of size, their intensities relative to the bright exciton line show a clear increase with confinement (Figure S9) in accordance with recent reports~\cite{choLuminescenceFineStructures2021,choExcitonPhononTrion2022a}. The Huang Rhys factors that characterise the exciton-phonon coupling strength increase from $\sim$0.05 to 0.2 in the range of sizes studied here and are found to be smaller than in the hybrid perovskite counterparts~\cite{choLuminescenceFineStructures2021}. %
The energies of the modes identified here energetically close to the widely observed transverse optical modes in Raman measurements interpreted in terms of bulk modes~\cite{yaffeLocalPolarFluctuations2017,iaruFrohlichInteractionDominated2021}. However, direct comparison between the optical phonon modes observed in single NC spectroscopy and those observed in Raman measurements should be done carefully as different selection rules apply. %
In fact, similar signatures were identified based on DFT calculations of bulk orthorhombic CsPbBr\textsubscript{3} as longitudinal optical phonon modes of the orthorhombic perovskite lattice~\cite{choExcitonPhononTrion2022a}. These modes were found to be mainly related to the lead-halide cage, with rotations of the PbBr\textsubscript{6} octahedra and stretching of the Pb-Br bonds, while the inorganic cation also contributes to modes 1 and 3~\cite{choExcitonPhononTrion2022a}. %

In conclusion, our study thoroughly examines the spectral properties of single CsPbBr\textsubscript{3} NCs over a wide range of sizes and provides critical insights into the band-edge exciton fine structure, exciton-complexes and optical phonon replica that feed the scientific debate and will help refine theoretical models to converge towards a realistic description of these nascent materials.

The polarisation properties of the bright triplet exciton were analysed and the diversity of spectra is interpreted as arising from a combination of the orientation of the emitting dipoles and the population of the emitting states. The wide variety of bright triplet emission spectra and their polar diagrams can be reproduced. 
Bright-bright energy splittings were also recorded and reveal a size-dependence consistent with a pure exchange model, i.e. without the Rashba effect. 
However, because of the large number of effects involved in setting the exciton fine structure, namely crystal symmetry, NC morphology and dielectric environment, disentangling their respective contributions remains a challenge that may not be achieved solely by optical means. To that end, the correlation of optical studies with electron microscopy on the same single NCs would help to account for the influence of NC shape anisotropy. %
Our work also revealed the size-dependence of the spectral signatures of the trion and biexciton which are in good agreement with a simple quantum confinement model. %
For each emitting state, we also identified up to four phonon replica for the exciton, while the weaker emitting trion and biexciton reveal up to three replica with energy shifts matching those of the exciton. The absence of a size-dependence for phonon energies indicates that they are related to bulk optical phonon modes of the orthorhombic perovskite lattice rather than acoustic or surface optical modes.

In order to fully realise the potential of LHP NCs as quantum light emitters, significant developments must be achieved. In particular, for single photon emission, the control of the interaction of the emitter with its dielectric and electromagnetic environment is paramount. %
To minimise the dephasing rate, i.e. the interaction of the emitter with its dielectric environment, more stable surface ligands or core-shell structures could be investigated as for established wurtzite and zinc-blende NCs~\cite{efrosNanocrystalQuantumDots2021}. %

\section*{Supporting information}

Experimental methods, size-energy correspondence, single NC spectrum model, additional spectral information, Huang Rhys factors

\section*{Acknowledgements}

This work was supported by the French National Research Agency (ANR) through the project IPER-Nano2 (ANR-18-CE30-0023). M.-R.A. acknowledges PhD funding support from the Agency for Science, Technology \& Research (A*STAR) through the Singapore International Graduate Award (SINGA, SING-2018-02-0079). Q. X. gratefully acknowledges funding support from the National Natural Science Foundation of China (grant No. 12020101003, and 12250710126), and strong support from the State Key Laboratory of Low-Dimensional Quantum Physics at Tsinghua University. The authors thank Roland Schmidt from Hitachi and Adrian Sandu from Thermo Fisher for their help with the electron microscopy images. 

\def\bibsection{\section*{\refname}}
\bibliography{main}



\end{document}


\title{Supplementary information for:\texorpdfstring{\\}{ } Spectral fingerprint of quantum confinement in single \texorpdfstring{CsPbBr\textsubscript{3}}{CsPbBr3} nanocrystals}%

\author{Mohamed-Raouf \surname{Amara}}
 
\affiliation{%
 Laboratoire de Physique de l'\'Ecole Normale Sup\'erieure, ENS, Université PSL, CNRS, Sorbonne Universit\'e, Universit\'e Paris-Cité, F-75005 Paris, France
}%
\affiliation{%
 Division of Physics and Applied Physics, School of Physical and Mathematical Sciences, Nanyang Technological University, 637371, Singapore
}%

\author{Zakaria \surname{Said}}
\affiliation{%
 Laboratoire de Physique de l'\'Ecole Normale Sup\'erieure, ENS, Université PSL, CNRS, Sorbonne Universit\'e, Universit\'e Paris-Cité, F-75005 Paris, France
}%
\author{Caixia \surname{Huo}}%
\affiliation{%
 Division of Physics and Applied Physics, School of Physical and Mathematical Sciences, Nanyang Technological University, 637371, Singapore
}%
\author{Aurélie \surname{Pierret}}
\author{Christophe Voisin}
\affiliation{%
 Laboratoire de Physique de l'\'Ecole Normale Sup\'erieure, ENS, Université PSL, CNRS, Sorbonne Universit\'e, Universit\'e Paris-Cité, F-75005 Paris, France
}%
\author{Weibo \surname{Gao}}
\affiliation{%
 Division of Physics and Applied Physics, School of Physical and Mathematical Sciences, Nanyang Technological University, 637371, Singapore
}%
\author{Qihua \surname{Xiong}}
\altaffiliation{%
Current affiliation:
State Key Laboratory of Low-Dimensional Quantum Physics, Department of Physics, Tsinghua University, Beijing 100084, China.
}%
\altaffiliation{%
Frontier Science Center for Quantum Information, Beijing 100084, P.R. China
}%
\altaffiliation{%
Collaborative Innovation Center of Quantum Matter, Beijing 100084, P.R. China}%
\altaffiliation{%
Beijing Academy of Quantum Information Sciences, Beijing 100193, P.R. China}
\affiliation{%
 Division of Physics and Applied Physics, School of Physical and Mathematical Sciences, Nanyang Technological University, 637371, Singapore
}%
\author{Carole \surname{Diederichs}}
\email{Corresponding author: carole.diederichs@phys.ens.fr}
\affiliation{%
 Laboratoire de Physique de l'\'Ecole Normale Sup\'erieure, ENS, Université PSL, CNRS, Sorbonne Universit\'e, Universit\'e Paris-Cité, F-75005 Paris, France
}%
\affiliation{%
 Institut Universitaire de France (IUF), 75231 Paris, France
}%

\maketitle

\tableofcontents
\setcounter{section}{0}
\renewcommand{\thesection}{S\Roman{section}}
\renewcommand{\thesubsection}{\alph{subsection}}
\setcounter{figure}{0} 
\renewcommand\thefigure{S\arabic{figure}}
\renewcommand\thetable{S\arabic{table}}
\renewcommand*{\citenumfont}[1]{S#1}
\renewcommand*{\bibnumfmt}[1]{[S#1]}
\section{Methods}

\subsection{Synthesis and sample preparation}

CsPbBr\textsubscript{3} NCs were synthesised following a previously published synthesis~\cite{beckerBrightTripletExcitons2018} where the injection temperature was adjusted to obtain NCs of varying edge lengths. Four different batches, labelled A to D, were synthesised in this work.

Samples for single NC spectroscopy were obtained by diluting the obtained solutions in a 3\%-w.r.w. polystyrene in toluene solution. Typical dilution levels used for single NC spectroscopy are 1:3000. The diluted solutions were then spin-coated at 3000 RPM for 30s and the samples were kept under inert atmosphere (or primary vacuum) before cooling down to cryogenic temperature. With this procedure, we obtain polymer layer with thicknesses in the order of \SI{100}{\nano\meter} within which nanocrystals are randomly oriented.

\subsection{Cryogenic-temperature optical spectroscopy}

All optical measurements were performed with a home-built scanning confocal microscope.
Excitation was provided either by a continuous-wave laser diode emitting at \SI{457}{\nano\meter} or a frequency-doubled Ti:Sa laser tuned to $\sim$~\SI{450}{\nano\meter}. Excitation was focused on the sample by a microscope objective with a 0.7 numerical aperture and the emission was collected through the same objective. The emission is isolated from the excitation using a set of 90/10 (R/T) beamsplitter and long-pass filters, dispersed using a grating monochromator and detected with a CCD camera with a resolution close to \SI{200}{\micro\electronvolt}. %

\subsection{Size-energy correspondence}\label{sec:sizener}

\begin{figure*}
    \centering
    \hspace{10pt}\includegraphics[scale=0.9]{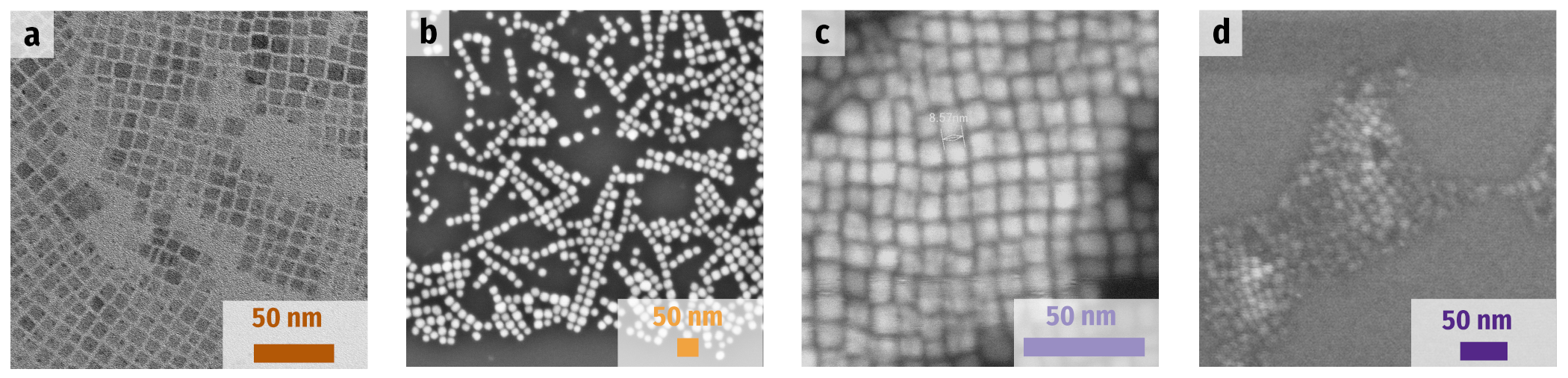}
\caption{Sample of transmission electron microscopy (A) and scanning electron microscopy (B, C, D) images used for the determination of the size and aspect ratio of NCs. Details in~\cite{Note1}.}
\label{fig:em_sample}
\end{figure*}

\begin{figure*}
\centering
\includegraphics[]{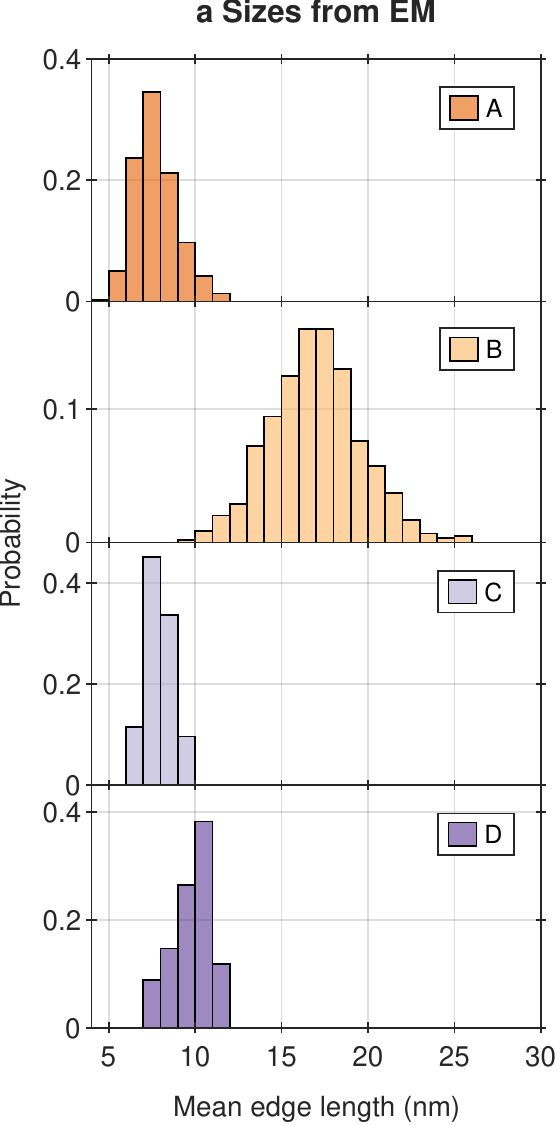}\hspace{10pt}\includegraphics[]{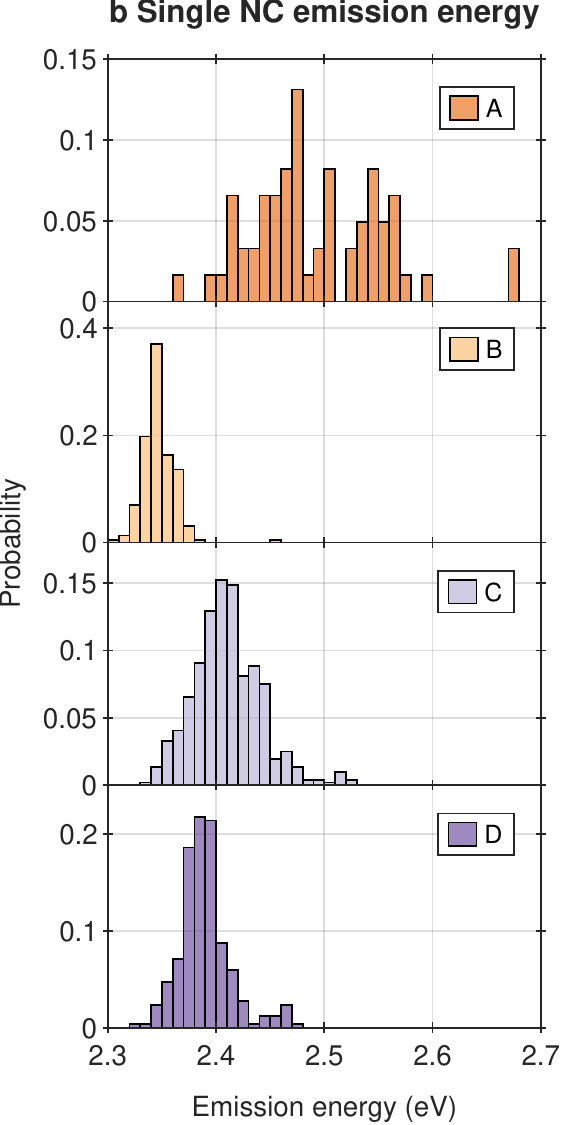}\hspace{10pt}\includegraphics[]{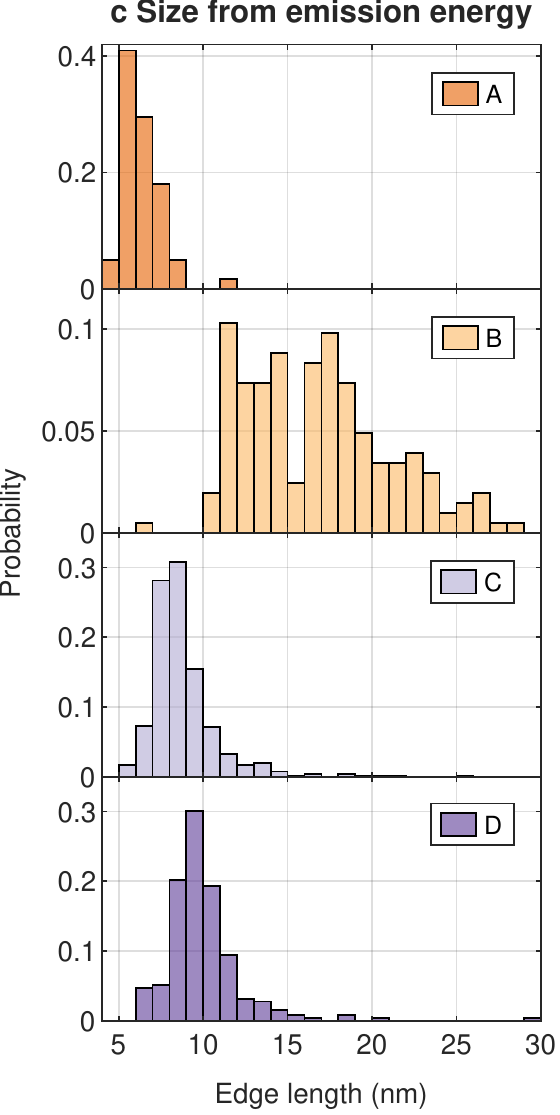}
\caption{Relationship between the NC size and emission energy for each synthesis batch. (a) Size distribution obtained from electron microscopy; averaged length between the two in-plane dimensions. (b) Distribution of single NC emission energies obtained from single NC PL. (c) Size distribution obtained by converting the single NC emission energies in (b) to sizes using the relationship determined in Figure S2. The size-energy correspondence is discussed in~\ref{sec:sizener}.}
\label{fig:size_energy}
\end{figure*}

\begin{figure}
    \centering
    \includegraphics[width=\textwidth]{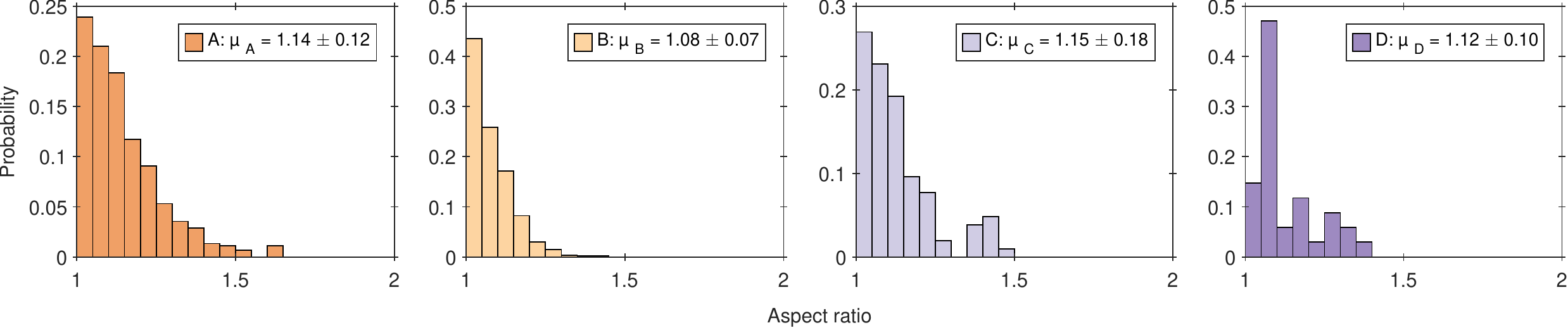}
    \caption{Aspect ratio determined from electron microscopy for each synthesis batch.}
    \label{fig:AR_from_em}
\end{figure}

Using either transmission or scanning electron microscopy (EM)~\footnote{Transmission electron microscopy images for synthesis A were acquired on a Jeol JEM. Scanning electron microscopy images for synthesis batches B, C and D were acquired respectively on a Thermo Fisher Apreo 2, Hitachi Regulus and Raith.}, we determined the distribution of NCs' in-plane dimensions. Sample images and the mean edge lengths recovered for each synthesis batch (A, B, C and D) are shown respectively in~\autoref{fig:em_sample} and~\autoref{fig:size_energy}a. The distribution of sizes for each synthesis are centered respectively at: A \SI{7.7\pm1.2}{\nano\meter}, B \SI{17.0\pm2.6}{\nano\meter}, C \SI{8.0\pm0.8}{\nano\meter}, D \SI{9.80\pm1.1}{\nano\meter}. We also record the aspect ratio deduced from the EM images and find mean aspect ratios close to 1.1 as shown in~\autoref{fig:AR_from_em}.

The distribution of single NC emission energies at cryogenic temperature are summarised for each synthesis batch in~\autoref{fig:size_energy}b. We use the distributions of mean edge lengths in \autoref{fig:size_energy}a and emission energies in \autoref{fig:size_energy}b to determine a size-energy correspondence as explained hereafter.

\begin{SCfigure}[5][hptb]
    \centering
    \includegraphics[width=0.4\linewidth]{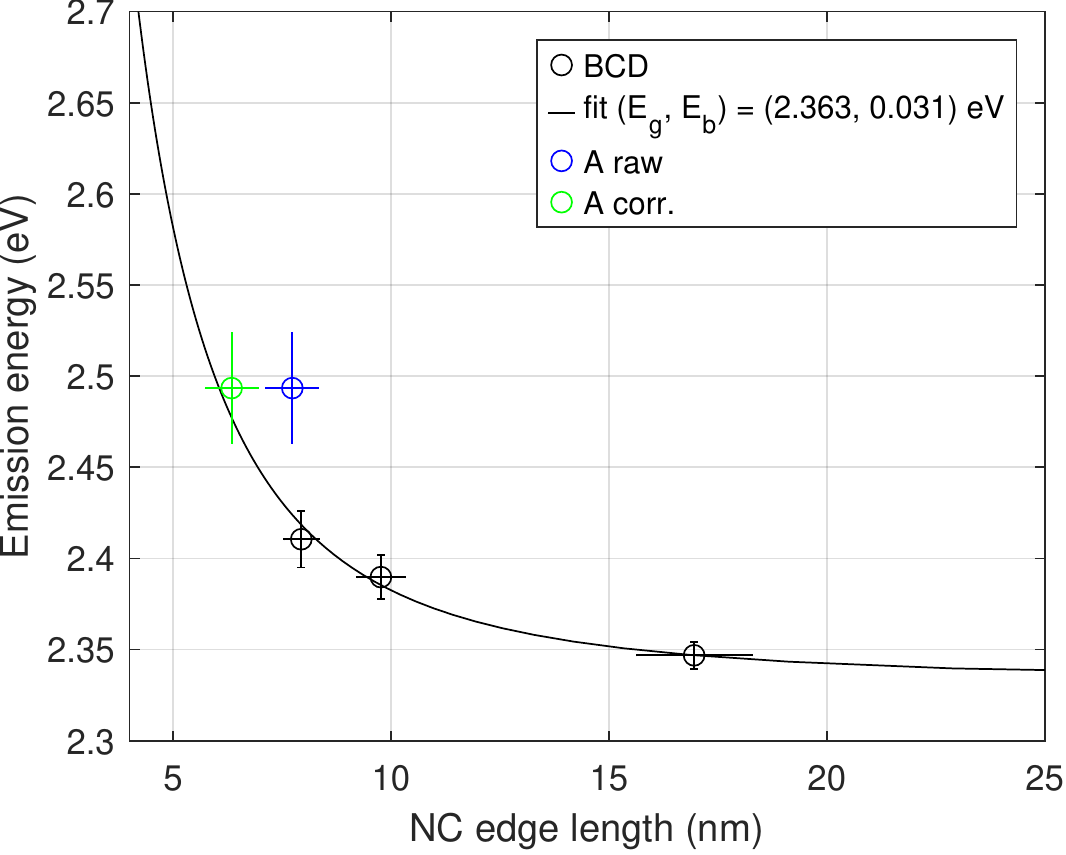}
    \caption{Correspondence between the distribution of single NC emission energies obtained from single NC PL and the distribution of NC edge lengths obtained from electron microscopy. Data points correspond to the mean edge lengths and emission energies recorded for each synthesis batch. Error bars display the standard deviation of each underlying distribution. The best fit is obtained using eq.~\eqref{eq:EofL} with $E_1 = \SI{2.363}{\electronvolt}$ and $E_2 = \SI{31}{\milli\electronvolt}$. Data points in blue, resp. green, correspond to the raw, resp. corrected, edge lengths recorded for synthesis A as explained hereafter.}
    \label{fig:corresp}
\end{SCfigure}

In~\autoref{fig:corresp}, we show as black points the average single NC emission energy as a function of the corresponding average NC edge length for synthesis batches B-D where the vertical and horizontal bars represent respectively the dispersions in emission energies and NC edge lengths.
The relationship between the NC size and the single NC PL energy is approximated using an expression for the exciton energy in the weak to intermediate confinement regime proposed by~\cite{sercelQuasicubicModelMetal2019}
\begin{equation}
    E_x \sim E_g + E_b \left[ \frac{3\pi^2}{(L/a_x)^2} - \sqrt{1+\left( \frac{2\times 3.05}{L/a_x}\right)^2}\right] \label{eq:EofL}
\end{equation}
where $E_g$ is the bandgap, $E_b$ the exciton binding energy, $a_x=\SI{3.05}{\nano\meter}$ the Bohr exciton radius~\cite{yangImpactHalideCage2017} and $L$ the NC edge length. Here, the two parameters $E_g$ and $E_b$ may not be directly related to the bulk bandgap and the exciton binding energy and are hereafter relabelled as $E_1$ and $E_2$.
Using this relationship and focusing only on synthesis batches B-D (black points), we find energies $E_1 = \SI{2.363}{\electronvolt}$ and $E_2 = \SI{31}{\milli\electronvolt}$. Extrapolating the law for larger NCs, we find a bulk bandgap at $\sim$\SI{2.34}{\electronvolt}, close to the reported bulk bandgap at liquid helium temperatures~\cite{yangImpactHalideCage2017}.

Let us note that for synthesis A (the smallest NCs), no direct comparison between the size distribution (obtained from EM) and the emission energy distribution (obtained from single NC PL) is possible. In fact, a significant experimental bias was introduced by focusing on the higher energy emitting (i.e. smallest) NCs of the sample, above $\sim$\SI{2.45}{\electronvolt}. Therefore, the emission energy distribution should not reflect the distribution of sizes, but rather corresponds to the smallest NCs of the distribution determined from EM data. The data point (in blue in~\autoref{fig:corresp}) is thus not used to establish the size-energy correspondence. 

Once a correspondence is established, we can however verify that the sizes obtained from the distribution of emission energies is coherent with the size distribution obtained from EM once the systematic bias is taken into account for A. With a selection threshold at \SI{2.45}{\electronvolt} corresponding to a size of \SI{7.0}{\nano\meter}, we truncate the EM size distribution and only keep sizes below \SI{7.0}{\nano\meter}. Doing so, the mean of the truncated size distribution becomes \SI{6.34}{\nano\meter} as shown in~\autoref{fig:corresp} in green which is consistent with the law deduced from the three other data sets and the known experimental selection.

\section{Polarisation properties of single NCs}

\subsection{Polarisation of three randomly oriented orthogonal dipoles}

Polarisation-resolved measurements are performed using a rotating half-waveplate placed before a polariser with its axis aligned with the monochromator grating preferential direction. Doing so, the strong polarisation dependence of the monochromator grating response does not play on the recovered signal. The half-waveplate is rotated and a series of spectra are acquired at each angular position. 

Doing so, we can retrieve a spectrum with genuine relative intensities by summing all polarisations and the polarisation direction of each emission line is recovered by fitting each spectrum with a set of lorentzian distributions. Figure 2 of the main text shows results of this procedure, where the spectrum is the unpolarised summed spectrum and the polar inset shows the intensity of each emission line as a function of the polarisation angle.
The emission intensity of each bright exciton is obtained by considering the emission of three orthogonal dipoles along $x$, $y$ and $z$ in a particular direction $\vec{u}$~\cite{beckerBrightTripletExcitons2018}. The polarisation of     the emission is measured in a plane perpendicular to $\vec{u}$ with a unit vector $\vec{e}_\alpha$ defined up to a rotation angle $\alpha$ in this plane.
The emission intensity detected from each dipole for a given polarisation angle is thus proportional to $\left|\vec{e}_\alpha\cdot\vec{x}\right|^2$, $\left|\vec{e}_\alpha\cdot\vec{y}\right|^2$, $\left|\vec{e}_\alpha\cdot\vec{z}\right|^2$. 
The emission intensity of each dipole is simulated for several angles to recover the polarisation dependence of the emission and construct a spectrum by summing over all angles. %
To relate these results to the experimental polarisation measurements, we further consider the relative population in the states corresponding to the emitting dipoles according to Boltzmann occupation factors, which directly depend on the temperature and on the energy splitting between the states. 
All in all, simulated spectra are thus compared to experimental data by setting the energy splittings and linewidths and allowing for the orientation and experiment temperature to vary. Doing so, we can find orientations that match experimental data with realistic experiment temperatures.

\begin{figure}[htbp]
    \centering
\includegraphics[width=\linewidth]{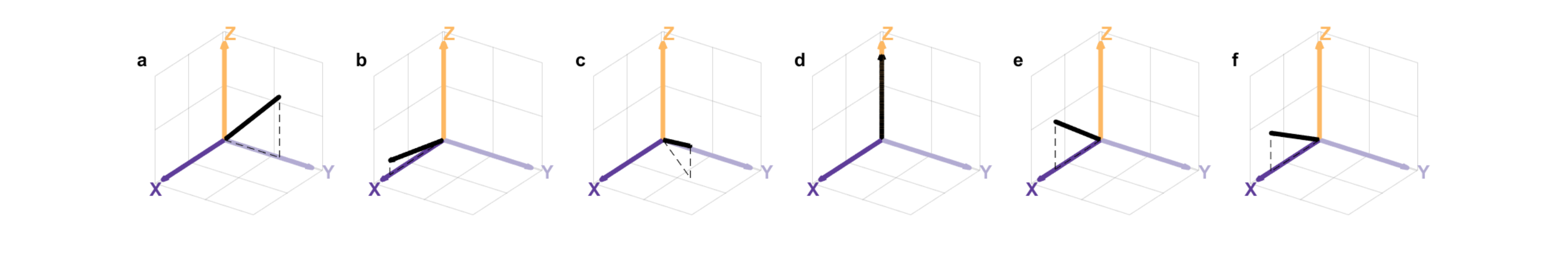}

\hspace{-45pt}\begin{tabular*}{0.85\textwidth}{l@{\extracolsep{\fill}}cccccc}
     $\Theta$ & 45 & 80.5 & 68.2 & 0  & 56.3 & 63.4 \\
     $\varPhi$ & 90 & 0 & 53.1 & 270 & 0 & 0
\end{tabular*}        
        \caption{Orientation of the observation axis (in black) with respect to the emitting dipoles (in colours) for each spectrum shown in Figure 2 of the main text.}
    \label{fig:myorientations}
\end{figure}

The observation directions of the emitting dipoles for the spectra in Figure 2 of the main text are summarised in \autoref{fig:myorientations}.

\subsection{Crystallographic axes vs. NC morphological axes}

As discussed in the main text, the orientation of the crystallographic axes with respect to the NCs morphological axes remains unclear in nanocubes, in contrast to nanoplatelets. %
In fact, a recent study based on transient absorption revealed a size- and temperature-dependent bright-triplet exciton energy splitting, even in the absence of shape anisotropy~\cite{hanLatticeDistortionInducing2022}. This splitting is attributed to an avoided crossing between excitons in LHP NCs because the NCs facets are aligned with pseudo-cubic \{100\} planes rather than the orthorhombic crystal planes. If such observations indeed hold true, only one of the three bright triplet transition dipole would be aligned with a NC edge while the other two match the orthorhombic crystal axes corresponding to diagonals of a NC facet rather than its edges.

\begin{figure*}[htbp]
\includegraphics[]{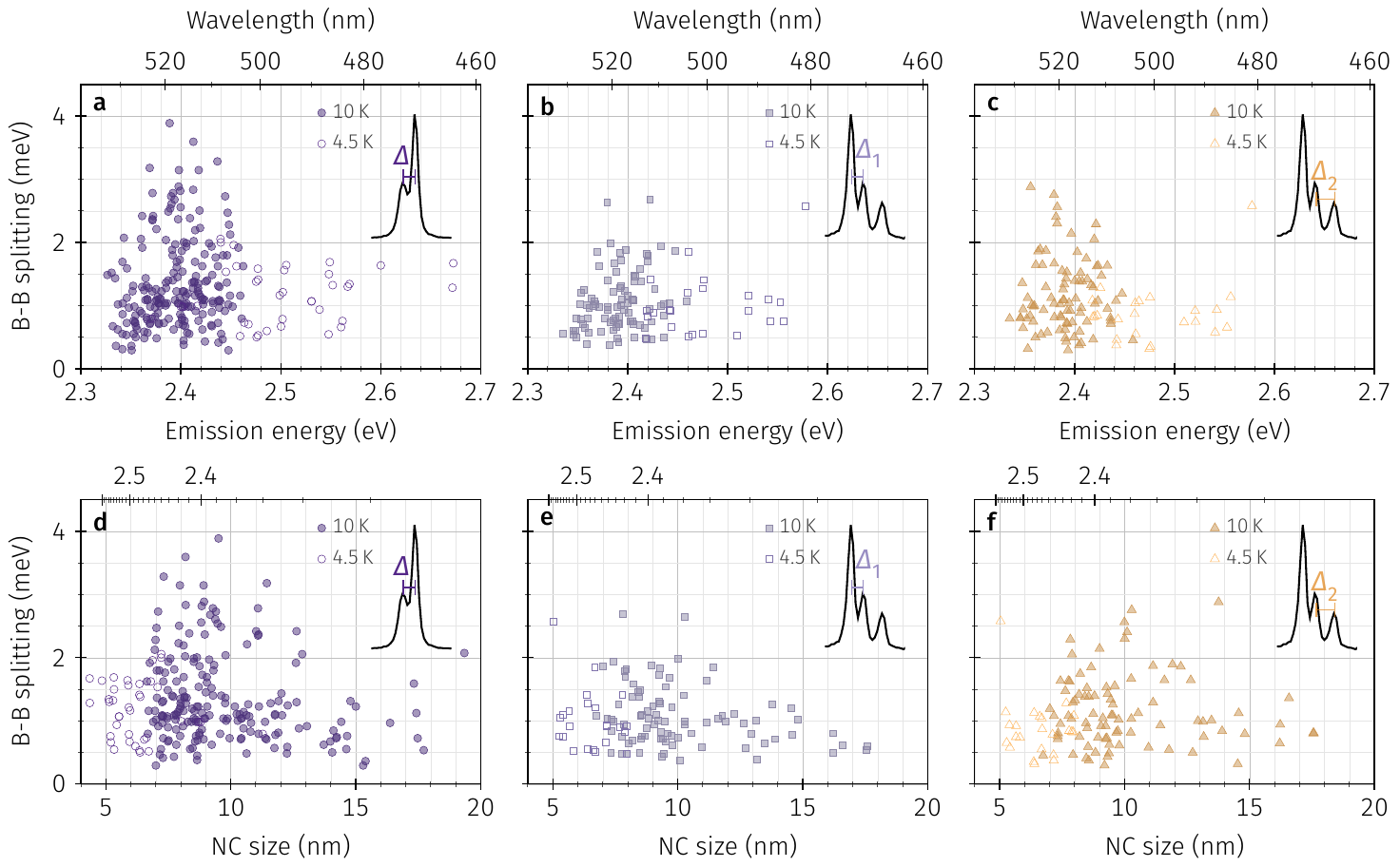}%
\caption{Bright-bright exciton energy splittings in single CsPbBr$_3$ NCs. Filled (empty) markers correspond to data acquired at \SI{10}{\kelvin} (\SI{4.5}{\kelvin}). }
\label{fig:splittings_full}
\end{figure*}

\section{Additional spectral information}

\subsection{Bright triplet splittings}

\autoref{fig:splittings_full} shows the recorded bright triplet energy splittings as a function of both the single NC emission energy (\autoref{fig:splittings_full}a-c) and the deduced NC size (\autoref{fig:splittings_full}d-f).

As explained in the figure legend, filled (empty) markers correspond to experiment data acquired at \SI{10}{\kelvin} (\SI{4.5}{\kelvin}). The main difference between the two datasets is expected to arise from the thermal population of the bright triplet exciton states. For two-peak spectra (\autoref{fig:splittings_full}a,d), the two peaks resolved can possibly correspond to any pair within the bright triplet (1-2, 2-3 or 1-3). Therefore, at the lowest temperatures investigated, we can expect that the proportion of two-peak spectra corresponding to the largest energy splittings (1-3) (i.e. splittings $\geq$~\SI{2}{\milli\electronvolt}), is minimised. This is consistent with the observation of two-peak energy splittings showing smaller average values at \SI{4.5}{\kelvin} than at \SI{10}{\kelvin} for similarly-sized NCs and for smaller NCs despite the expected increase of energy splittings for smaller NCs.

\subsection{Charged exciton, trion}

\begin{figure*}[htbp]
    \centering
\includegraphics{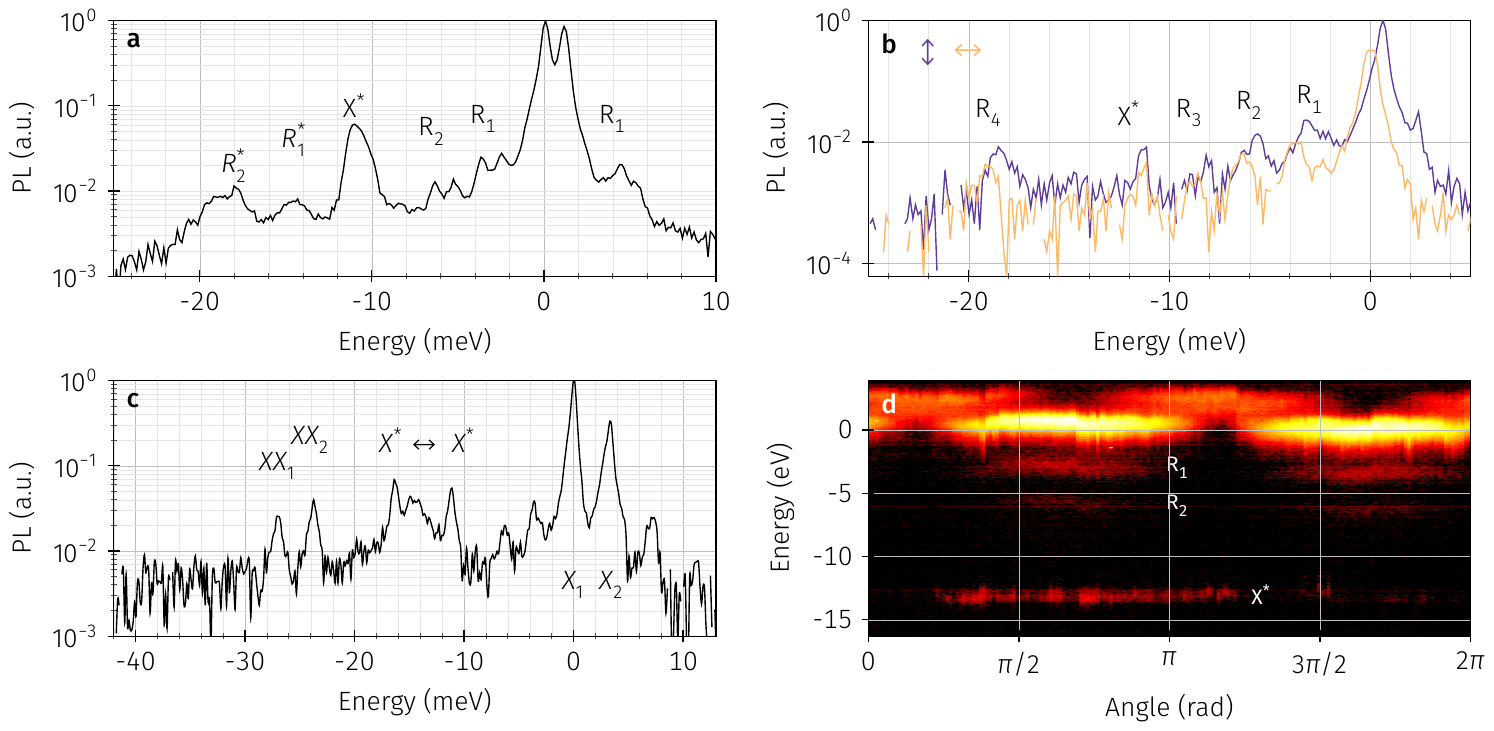}
    \caption{Additional spectra. (a) Single NC spectrum highlighting the absence of a trion fine structure and the presence of optical phonon replica for each bright triplet exciton peak. (b) Single NC spectra acquired at orthogonal polarisations showing the fine structure of the bright triplet exciton and the related optical phonon replica while the trion emission reveals no fine structure. (c) Single NC spectrum highlighting the fine structure of the biexciton which mirrors that of the bright triplet exciton. (d) Series of PL spectra acquired as a function of the detection polarisation highlighting the polarisation of the first two optical phonon replica related to the most intense bright exciton peak while the trion emission is intermittent and shows no clear polarisation.}
    \label{fig:suptrionbi}²
\end{figure*}

\autoref{fig:suptrionbi} shows additional information related to the spectral features identified in this work. 

\autoref{fig:suptrionbi}a highlights the fine structure of the bright triplet exciton and that of the associated optical phonon replica while the trion emission and its replica do not show a fine structure. In \autoref{fig:suptrionbi}b, we resolve the fine structure of the biexciton with an energy splitting matching that of the bright exciton. The trion emission in \autoref{fig:suptrionbi}b shows a complicated spectrum related to spectral wandering during the acquisition and overlap with the second set of optical phonon replica related to the bright exciton doublet. 

In \autoref{fig:suptrionbi}c, we show spectra of a single NC exhibiting two peaks attributed to the bright exciton. Each spectrum is acquired with a polarisation axis parallel to one of the bright exciton polarisation axis and reveals the phonon replica associated with this bright exciton state. Notably, the trion emission does not show any polarisation dependence. This is further illustrated in~\autoref{fig:suptrionbi}d which shows the emission spectrum of a single NC as a function of the detected polarisation angle. While optical phonon replica associated with the bright exciton show the same polarisation as the bright exciton, the trion emission does not follow the same polarisation dependence and exhibits notable spectral and intensity fluctuations.  

\subsection{Optical phonon replica}
    
\begin{figure*}[htbp]
\centering
\includegraphics{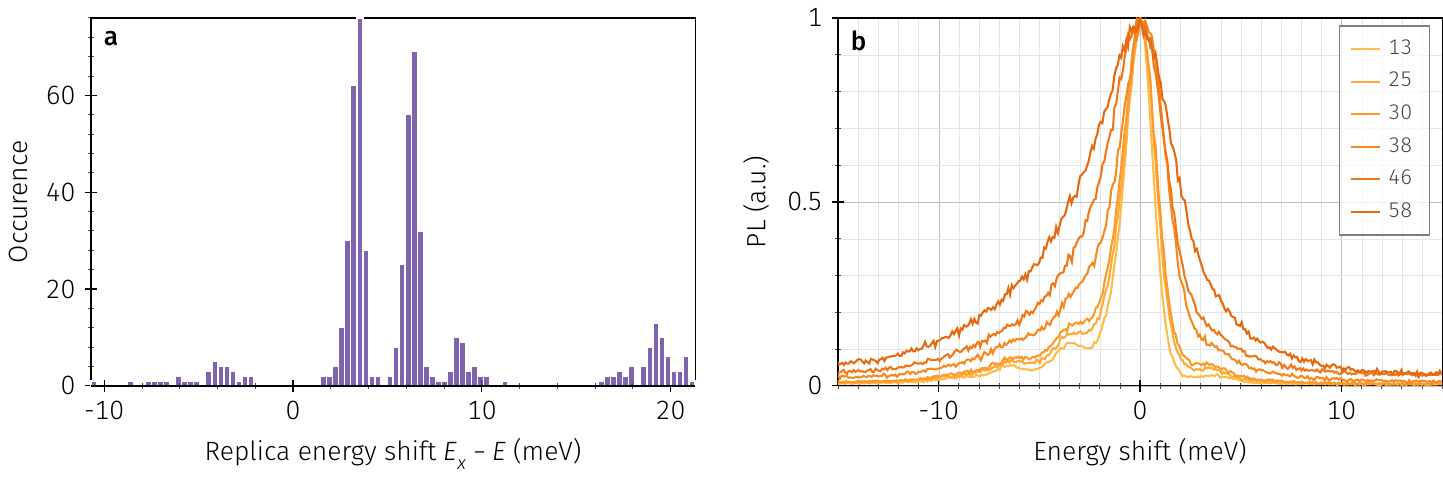}
\caption{Optical phonon replica. (a) Distribution of Stokes and anti-Stokes optical phonon replica recorded in single NCs PL spectrum. (b) Spectrum of a single NC as a function of temperature.}
\label{fig:stokesantistokes}
\end{figure*}

Optical phonon replica detected in the low-temperature PL of single NCs are summarised in~\autoref{fig:stokesantistokes}a where both Stokes and anti-Stokes replica are shown. %
\autoref{fig:stokesantistokes}b shows the temperature dependence of a single NC spectrum. As temperature is increased, the fraction of the emission in the phonon replica increases together with a general increase of the emission linewidth. These two effects result in a broadening of the emission which presents an asymmetric lineshape at higher temperature, as previously observed e.g. for acoustic phonons in epitaxial quantum dots~\cite{faveroAcousticPhononSidebands2003}.

\begin{figure*}[htbp]
\centering
\includegraphics{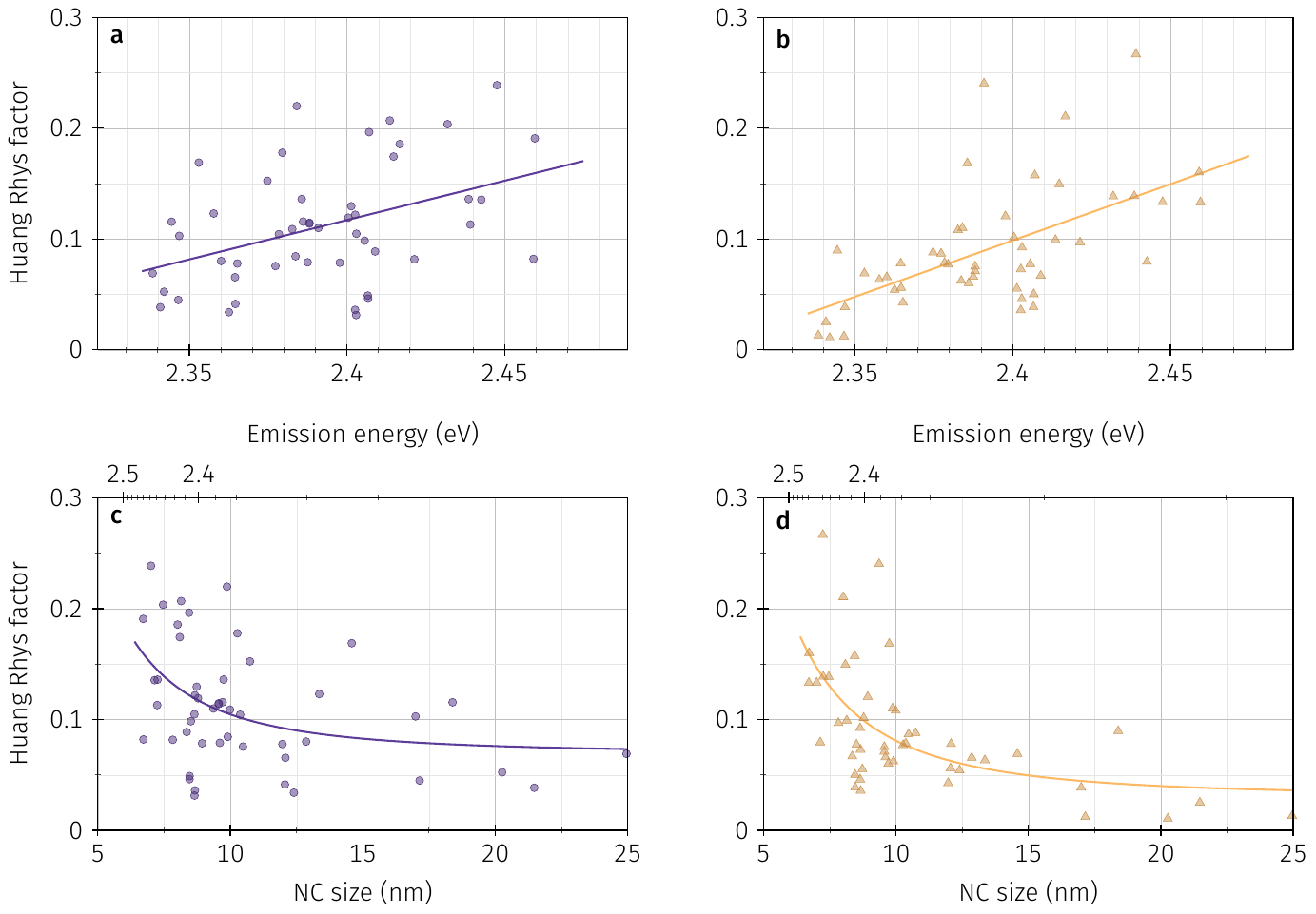}
\caption{Huang Rhys factors. (a,c) Huang Rhys factors of the first phonon replica shown as a function of (a) emission energy and (c) NC size. (b,d) Huang Rhys factors of the second phonon replica shown as a function of (b) emission energy and (d) NC size.}
\label{fig:huangrhysfactor}
\end{figure*}

The relative intensities of the two lowest-energy optical phonon replicas are recorded as a function of the emission energy of the bright exciton state. Assuming a contribution of the first replica of each of the two lowest energy phonon modes identified, we can determine the corresponding Huang Rhys factors as shown in~\autoref{fig:huangrhysfactor} as a function of the emission energy and corresponding NC size.

\clearpage
\bibliography{main}